\documentclass[12pt]{iopart}

\usepackage{epsfig}

\newcommand{\dd}{\mathrm{d}}  
\newcommand{\ttr}{\mathrm{Tr}} 
\newcommand{\partialbar}{\partial\kern -5pt/}
\newcommand{\Dbar}{\mathcal{D}\kern -6pt/}

\newcommand{\Mpl}{M_{\rm Pl}}

\newcommand{\sW}{\mathcal{W}}
\newcommand{\sN}{\mathcal{N}}
\newcommand{\sK}{\mathcal{K}}
\newcommand{\sL}{\mathcal{L}}
\newcommand{\sD}{\mathcal{D}}
\newcommand{\sP}{\mathcal{P}}
\def\lsim{\hbox{ \raise.35ex\rlap{$<$}\lower.6ex\hbox{$\sim$}\ }}
\def\gsim{\hbox{ \raise.35ex\rlap{$>$}\lower.6ex\hbox{$\sim$}\ }}

\begin{document}

\title{D-term inflation in non-minimal supergravity}
\author{Jonathan Rocher$^\dagger$ and Mairi Sakellariadou$^\star$}

\address{$^\dagger$ Theory Group, Department of Physics, University of
Texas Austin, Texas 78712,
USA}
\address{$^\star$ Department of Physics, King's College, University
of London, Strand, London WC2R 2LS, United Kingdom}
\ead{$^\dagger$rocher@physics.utexas.edu, 
$^\star$Mairi.Sakellariadou@kcl.ac.uk}

\begin{abstract}
D-term inflation is one of the most interesting and versatile models
of inflation. It is possible to implement naturally D-term inflation
within high energy physics, as for example SUSY GUTs, SUGRA, or string
theories. D-term inflation avoids the $\eta$-problem, while in its
standard form it always ends with the formation of cosmic
strings. Given the recent three-year WMAP data on the cosmic microwave
background temperature anisotropies, we examine whether D-term
inflation can be successfully implemented in non-minimal supergravity
theories.  We show that for all our choices of  K\"ahler potential,
there exists a parameter space for which the predictions of D-term
inflation are in agreement with the measurements. The cosmic string
contribution on the measured temperature anisotropies is always
dominant, unless the superpotential coupling constant is fine tuned; a
result already obtained for D-term inflation within minimal
supergravity. In conclusion, cosmic strings and their r\^ole in
the angular power spectrum cannot be easily hidden by just considering
a non-flat K\"ahler geometry.
\end{abstract}
\maketitle

\section{Introduction and motivations}

Inflation offers a simple solution to the shortcomings of the
standard hot big bang model. In addition, its predictions about
the initial density perturbations, leading to the observed
structure formation, are in a remarkable agreement with
measurements of the Cosmic Microwave Background (CMB) temperature
anisotropies.  Among the various inflationary models, one should
select the ones which lead to a better agreement with data, with
the additional requirement that such models should be naturally
built within a fundamental theoretical framework.  In spite of its
enormous success, inflation remains still a paradigm in search of
a model.

Despite the elegance of chaotic inflation~\cite{chaotic,lindebook},
this simple model faces a fine tuning problem. Consistency, between
predictions for the amplitude of CMB temperature anisotropies and
measurements, requires a tiny coupling constant. To avoid this
problem, Linde~\cite{inflahyblinde1,inflahyblinde2} has proposed
hybrid inflation, a model of inflation based on Einstein gravity, but
driven by a false vacuum. In this model, the inflaton field rolls
while another scalar field remains trapped in a false vacuum
state. The false vacuum becomes unstable when the magnitude of the
inflaton field falls below some critical value, leading to a phase
transition to the true vacuum. The energy density is dominated by the
false vacuum energy density so that the phase transition signals the
end of hybrid inflation. The phase transition at the end of inflation
leads to topological defect formation~\cite{copeland94}.

Theoretically motivated inflationary models can be built in a
context of Supersymmetry (SUSY) and Supergravity (SUGRA) theories.
$N=1$ supersymmetric models contain complex scalar fields, which
often have flat directions in their potential, thus offering
natural candidates for inflationary models. In this framework,
hybrid inflation (driven by the F-terms or the D-terms) are the
most standard models. Such inflationary models lead generically to
cosmic string formation at the end of the inflationary era. Cosmic
strings in supersymmetric theories may have new properties, as
compared to their non-supersymmetric counterparts; we do not
address this issue here.

A gauge symmetry can be broken spontaneously in $N=1$ globally
supersymmetric theories, either by adding F-terms to the
superpotential or, in the Abelian case, by introducing
Fayet-Iliopoulos (FI) D-terms.  The Higgs mechanism leads
generically~\cite{prd2003} to Abrikosov-Nielsen-Olesen (ANO)
strings. Depending whether they were formed at the end of F- or D-term
inflation they are called F-term or D-term strings, respectively.
F-term inflation is potentially plagued with the $\eta$-problem, while
D-term inflation avoids it. This problem arises from the presence of
large corrections (of the order of the Hubble parameter during
inflation) to the inflaton mass, which spoil the required flatness of
the inflaton potential. D-term inflation can be successfully
implemented in the framework of SUGRA, while in addition, it can be
easily accommodated within string theory models.

In the simplest models of D-term inflation within SUGRA, in which
the constant FI term gets compensated by a single complex scalar
field at the end of the inflationary era, the D-term strings
formed at the end of inflation are topologically stable, since
$\pi_1({\cal M})\neq I$, with ${\cal M}$ the vacuum manifold of
the broken U(1) symmetry.  Our study concerns such models.
However, they have been proposed~\cite{va,uad,toine1} models where
D-term strings can become unstable.  For example, one can
introduce additional matter multiplets so as to obtain a
non-trivial global symmetry such as SU(2), leading to a simply
connected vacuum manifold and the production of semi-local
strings. Alternatively, it has been suggested~\cite{toine1} that
the waterfall Higgs fields are non-trivially charged under some
other gauge symmetries $H$, such that the vacuum manifold,
$[H\times U(1)]/ U(1)$, is simply connected, leading to the
formation of semi-local strings.

D-term inflation requires the existence of a non-zero constant FI
term, which can be added to the Lagrangian only in the presence of a
U(1) gauge symmetry. This extra U(1) symmetry can be of a different
origin. Some models have been suggested for field-dependent FI terms,
arising in the presence of a chiral superfield $\Phi$ shifting under
U(1). The imaginary part of the scalar part of the chiral superfield
plays the r\^ole of an axion, and cancels the chiral anomaly by
shifting under the U(1) symmetry.  Such a U(1) symmetry is called {\sl
anomalous}, or {\sl pseudo-anomalous} since the total anomaly
vanishes. Here the FI term depends on the real part of the chiral
superfield. In supersymmetry, models with anomalous FI terms have been
developed~\cite{anomalous} within heterotic string theory.  In a
cosmological setup one has first to assume the stabilisation of the
chiral superfield $\Phi$; the r\^ole of $\Phi$ may be played by any
modulus (a dilaton or any volume modulus). Only if this assumption
holds the dilaton-dependent D-term can be considered as a constant FI
term. However, the issue of dilaton and moduli stabilisation in the
heterotic string theory is far from being resolved. It is still not
clear how to derive constant FI terms from string or M-theory and only
field-dependent D-terms have been identified. As we will discuss, in
absence of constant FI terms, local supersymmetry requires the
superpotential to be invariant under the U(1)-gauged symmetry.

In the context of theories with large extra dimensions, brane
inflation occurs in a similar way as hybrid inflation within
supergravity, leading to cosmic string-like objects.  In string
theories, D-brane $\bar{\rm D}$-anti-brane annihilation leads
generically to the production of lower dimensional D-branes, with D3-
and D1-branes, which are D-strings, being predominant~\cite{md,rmm} in
IIB string theories.  To illustrate brane inflation let us
consider~\cite{sen} a D$p$-${\bar{\rm D}}p$ system in IIB string
theory. Six of the spatial dimensions are compactified on a torus,
while the branes move relatively to each other in some directions. As
the two branes approach, the open string modes between the branes
develop a tachyon, thus an instability.  Brane
inflation~\cite{dvaliandtye} ends by a phase transition mediated by
open string tachyons. Since the tachyonic vacuum has a non-trivial
$\pi_1$ homotopy group, one concludes that there must exist stable
tachyonic string solutions with $p-2$ co-dimensions; they are stable
BPS (Bogomol'nyi-Prasad-Sommerfield) D$(p-2)$-branes. Since all
dimensions are compact these daughter branes are seen as
one-dimensional objects for a four-dimensional observer; they are the
D-strings.

The D-strings (D1-branes) have been identified~\cite{Dstr-Dterm}
in the low-energy supergravity with the D-term strings.  The
justification for this conjecture is that only D-term strings
remain BPS states in $N=1, d=4$ supergravity. In supergravity,
F-term strings break all the supersymmetries, whereas the D-term
strings preserve half of it.

An interesting and successful brane inflationary model is the D3/D7
one~\cite{D3/D7}, which has also an effective description as a D-term
inflationary model.  The flat direction of the inflaton potential is
associated with the shift symmetry, which protects the inflaton field
from acquiring a large mass that would spoil the required flatness of
the potential.  In its original version this model leads to the
formation of topologically stable ANO BPS strings.  In a later
developed version~\cite{D3/D7new} of the D3/D7 model, where in terms
of an effective gauge theory the model has a local U(1) gauge symmetry
and a global SU(2) symmetry, semi-local strings are formed. Such
strings can unwind without any cost of potential energy. In what
follows we concentrate on inflationary models leading to the formation
of topologically stable strings.

Strings formed at the end of an inflationary era,
contribute~\cite{rachel97,kl,lr1,lr2,chm,bw,bprs} in the spectrum
of temperature anisotropies; their contribution is heavily
constrained from CMB data. Compatibility between CMB measurements
and theoretical predictions constrain~\cite{jcap2005,prl2005} the
parameters space (mass scales and couplings) of the inflationary
models.  These constraints have been
obtained~\cite{jcap2005,prl2005} for D-term inflation within
minimal SUGRA.  Here, we would like to investigate whether the
constraints on the parameters space are a result of our choice of
minimal supergravity, or whether they are a generic outcome of
D-term hybrid inflation. We will therefore examine D-term
inflation originated by different choices of the K\"ahler potential.
The main motivation being that a minimal K\"ahler potential can be
considered as a peculiar and unmotivated choice~\cite{copeland94}.

We base our study in a formulation of supergravity constructed from
superconformal theory, since the standard formulation may be
insufficient in the presence of constant Fayet-Iliopoulos
terms~\cite{toine1}.

To be able to constrain the parameters space of the models we
should know the power spectrum of a cosmic strings network and the
allowed upper limit on the cosmic string contribution to the
measured temperature anisotropy spectrum. The upper limit imposed
on the cosmic string contribution to the CMB data depends on the
numerical simulation employed in order to calculate the cosmic
string power spectrum. The upper limit found in the
literature~\cite{fraisse} is $7 \%$ or $11 \%$, depending on the
simulation, with $95 \%$ confidence level. There are some
uncertainties in these results due to the cosmic string evolution
codes they are based on\footnote{The upper limit of $7 \%$ was
found using the results of Ref.~\cite{bra}, based on the
velocity-dependent one-scale model for the string evolution. This
approach is certainly not the best. The upper limit of $11 \%$ was
found using the results of Ref.~\cite{pog}; however recently the
authors have corrected~\cite{pog2} a mistake in their code, which
has not been taken into account in Ref.~\cite{fraisse}. Note
however that the corrected version of the code \cite{pog2} seems
to give similar results. We believe that the recent publication
given in Ref.~\cite{marketal} is the best, at present, approach.}
and therefore in our calculations we use an average value for the
upper bound equal to $9\%$. Note that none of the existing
simulations take into account a non trivial microstructure of the
cosmic strings formed: however they have been shown to be
superconducting in supersymmetric abelian symmetry breaking
through F- or D-terms~\cite{davis}.

We plan the rest of the paper as follows: In Section II we address
hybrid D-term inflation, first within the standard formulation of
supergravity, and subsequently within the effective supergravity
theory built upon superconformal field theory. In the rest of the
paper we focus on the second approach. We briefly review the effective
supergravity formulation because of its consequences for D-term
inflation, often not taken into consideration. In Section III we
review D-term inflation in minimal supergravity. In Section IV we
discuss inflation in a supergravity theory with shift symmetry.  In
Section V we consider D-term inflation in supergravity models with
higher order terms, including all corrections up to order $M_{\rm
Pl}^{-2}$. We round up our conclusions in Section VI.

\section{Hybrid inflation and supergravity formulations}
We first review hybrid D-term inflation model in its standard
supergravity formulation.

\subsection{Standard formulation of supergravity}\label{sec:stdSUGRAformul}
Following the standard formulation of
supergravity~\cite{bailinlove,nilles}, the general SUGRA Lagrangian
for chiral superfields $\Phi_i$, and a vector superfield, depends on
three generic functions: the K\"ahler potential
$K(\Phi_i,\bar{\Phi}_i)$, the superpotential $W(\Phi_i)$, and the
kinetic function $f_{ab}(\Phi_i)$ for the vector multiplets. It can be
expressed as an integral over the Grassmann variables $\theta$ and
$\bar\theta$ (over superspace), as~\cite{bailinlove}:
\begin{eqnarray}
\mathcal{L}_{\rm SUGRA}=&&\int \dd^2\theta\dd^2\bar\theta \,
K(\Phi_i^\dagger e^{2gV},\Phi_i) + \int \dd^2\theta
[W(\Phi_i)+{\rm h.c.}]\nonumber\\
&& +\int \dd^2\theta [f_{ab}(\Phi_i)W_a^{\
\alpha} W_{\alpha b} +{\rm h.c.}]~, \label{lagrSUGRA}
\end{eqnarray}
where {\sl h.c.} stands for Hermitian conjugate.  It turns out that
the effective Lagrangian obtained from Eq.~(\ref{lagrSUGRA}) depends
only on one combination of the K\"ahler potential
$K(\Phi_i,\bar{\Phi}_i)$ and the superpotential $W(\Phi_i)$, the
following:
\begin{equation}
\label{kwcomb}
G(\Phi_i,\bar{\Phi}_i)= \frac{K(\Phi_i,\bar{\Phi}_i)}{\Mpl^2} +\ln
\frac{|W(\Phi_i)|^2}{\Mpl^6}~.
\end{equation}
A priori, the only restriction for the choice of the K\"ahler
potential $K$ and the superpotential $W$ is that $K$ must be a
real function and $W$ must be a holomorphic function.  However,
there exists a degeneracy in the choices of $K$ and $W$, while the
Lagrangian is invariant under the \emph{K\"ahler transformation}
\begin{eqnarray}\label{defkahltransf}
K(\Phi_i,\bar{\Phi}_i) &&\rightarrow K(\Phi_i,\bar{\Phi}_i) + h(\Phi_i)
+ h^\star(\bar{\Phi}_i) \nonumber\\ W(\Phi_i)&&\rightarrow e^{-h}W(\Phi_i)~.
\end{eqnarray}
Assuming that the Lagrangian must be invariant also under a gauge
symmetry, $W$ and $K$ must be invariant, at least up to a K\"ahler
transformation. The gauge kinetic function $f_{ab}(\Phi_i)$ must be
holomorphic and covariant under gauge symmetry.  Choosing for example
the \emph{minimal K\"ahler potential}
\begin{equation}
K_{\rm min}(\Phi_i,\bar{\Phi}_i)=\sum_i |\Phi_i|^2~,
\end{equation}
the kinetic terms for the scalar parts of the $\Phi_i$'s are
simply\footnote{We use the notations of Ref.~\cite{bailinlove} for
indices.}
\begin{equation}
K^i_j D_\mu \phi_i D^\mu \phi^{j^*} = D_\mu \phi_i D^\mu
\phi^{i^*}~.\nonumber
\end{equation}
In minimal SUGRA, we also set the gauge kinetic function $f_{ab}$
equal to $\delta_{ab}$.

Within this framework of supergravity, and assuming that there is a
gauge symmetry, the scalar potential $V$ for the scalar components
$\phi_i$ of the superfields $\Phi_i$ reads
\begin{equation}\label{formPotenScal}
V=\frac{e^G}{M_{\rm Pl}^4}\left[G_i(G^{-1})^i_jG^j-3 \right] +
\frac{1}{2}\left[{\rm Re} f_{ab}(\Phi_i)\right]^{-1}\sum_a g^2_a
D^2_a~;
\end{equation}
the first term in the \emph{r.h.s} of Eq.~(\ref{formPotenScal}) is
called \emph{F-term} and the second one \emph{D-term}. While the
F-term has in general positive and negative contributions, the D-term
is positive definite. $D_a$ is given by
\begin{equation}
D_a=\phi_i {(T_a)^i}_j \,K^j+\xi_a~,
\end{equation}
where the Fayet-Iliopoulos term, $\xi_a$, is non-zero only if the
gauge symmetry is Abelian; $g_a$ is the gauge coupling of the
symmetry generated by $T_a$. Assuming a constant FI term, D-term
potential may lead to de Sitter type of solutions, which is
particularly interested for building an inflationary model within
supergravity.

The standard D-term hybrid inflation model is based on the
superpotential~\cite{DSUGRA1,DSUGRA2}
\begin{equation}\label{superpoteninflaD}
W=\lambda S\Phi_+\Phi_-~,
\end{equation}
where $S, \Phi_+, \Phi_-$ are three chiral superfields and $\lambda$
is the superpotential coupling. This model assumes an invariance under
an Abelian gauge group $U(1)_\xi$, under which the three superfields,
$S, \Phi_+, \Phi_-$, have charges $0$, $+1$ and $-1$,
respectively. This model also assumes the existence of a constant FI
term\footnote{A supersymmetric description of the standard D-term
inflation is insufficient, the reason being that the inflaton field
reaches values of the order of the Planck mass, or above it, even if
one concentrates only around the last 60 e-folds of
inflation~\cite{prl2005,jcap2005}. The correct analysis is indeed in
the context of supergravity~\cite{prl2005,jcap2005}.}.

However, the above summarised supergravity formulation (called
hereafter \emph{standard}) is inappropriate to describe D-term
inflation~\cite{toine1}. Indeed, in D-term inflation the
superpotential vanishes at the unstable de Sitter vacuum, as it also
vanishes in the absolute Minkowski vacuum; anywhere else the
superpotential is non-zero. Thus, the standard formulation of
supergravity, where the Lagrangian depends on $K$ and $W$ only through
the combination given in Eq.~(\ref{kwcomb}) is inappropriate, since
the theory is ill defined at $W=0$. In conclusion, D-term inflation in
supergravity must be described with a non-singular formulation of
SUGRA when the superpotential vanishes.

\subsection{Effective supergravity from superconformal field theory}
\subsubsection{Superconformal Lagrangian}
Various formulations of effective supergravity can be constructed from
the superconformal field theory (see, Ref.~\cite{SCFT} for a recent
review). The main idea is, in a first stage, to build a Lagrangian
with full superconformal theory. In a second stage, the gauge
symmetries that are absent in Poincar\'e supergravity (e.g., local
dilatations, local chiral U(1)-symmetry and local S-supersymmetry) are
gauge fixed. Starting from this framework in order to derive the
effective supergravity theory, one can construct a non-singular theory
at $W=0$, where the action depends on all three functions, $K, W$ and
$f_{ab}$.

The Superconformal Field Theory (SCFT) is based on the $SU(2,2|1)$
symmetry. The superconformal Lagrangian contains three parts, each of
them being separately conformally invariant. For $n+1$ chiral
multiplets $X_I$ (with $(X_I)^\star\equiv X^I)$ and some vector
multiplets $\lambda^\alpha$ superconformally coupled to supergravity,
the superconformal Lagrangian reads~\cite{toine1}
\begin{equation}\label{lagrangSuperconf}
\mathcal{L}_{\rm SCFT}=\left[ \sN(X,X^\star)\right]_D + \left[\sW(X)
\right]_F + \left[f_{\alpha\beta}(X) \bar{\lambda}^\alpha
\lambda^\beta\right]_F~.
\end{equation}
The homogeneous function $\sN(X,X^\star)$ and the holomorphic
functions $\sW(X)$, $f_{\alpha\beta}(X)$  encode the K\"ahler
potential, the superpotential and gauge kinetic function,
respectively, once the extra gauge symmetries have been
gauge-fixed.

After constructing the Lagrangian, we fix the extra symmetries.
Fixing local dilatation, the number of chiral scalars is decreased by
one : there will be only $n$ physical scalar fields in
supergravity. Fixing S-supersymmetry makes a free fermion field to be
removed. One thus makes a transformation from the $n+1$ variables
$X_I$ to the \emph{conformon} scalar $Y$ and $n$ physical scalars
$z_i$, the $n$ chiral superfields of standard supergravity.  More
precisely,
\begin{equation}
X_I=Y x_I(z_i)~,
\end{equation}
where $x_I$ are a set of holomorphic functions. The K\"ahler potential
$\sK$ is related to the function $\sN(X,X^\star)$, appearing in the
superconformal Lagrangian, and the conformon superfield $Y$ through
\begin{equation}\label{definkahler}
\sK(z,z^\star)=-3\ln\left(-\frac{1}{3}\sN/YY^\star \right)~.
\end{equation}
Chiral and dilatation symmetry imply that the holomorphic function
$\sW$ is
\begin{equation}
\sW(Y,z)=Y^3 \Mpl^{-3}W(z)~.
\end{equation}

Apart the local $SU(2,2|1)$ symmetry, the Lagrangian may also have
some Yang-Mills gauge symmetries, which commute with local
superconformal symmetries. The superconformal functions $\sN, \sW$ are
invariant, while the superconformal function $f_{ab}$ is covariant
under the Yang-Mills gauge symmetries. The Yang-Mills transformations
of all chiral superfields in the superconformal action are
\begin{equation}
\delta_\alpha Y =Y r_\alpha(z)~~ \mbox {and}\quad \delta_\alpha z_i
=\eta_{\alpha i}(z)~,
\label{splitting}
\end{equation}
$r_\alpha(z)$ and $\eta_{\alpha i}(z)$ are $n+1$ holomorphic functions
for every symmetry. Note that the splitting given in
Eq.~(\ref{splitting}a) is not unique; the action invariance under
K\"ahler transformation, Eq.~(\ref{defkahltransf}), has its origin
exactly in this remark. The meaning of the value of $r_\alpha$ is the
transformation of the conformon field $Y$.

The invariance of $\sN$ leads to~\cite{toine1}
\begin{equation}\label{ralpha}
0=\sN\left\{r_\alpha(z)+r^\star_\alpha(z^\star)-\frac{1}{3}\left[\eta_{\alpha
i}
\partial^i\sK(z,z^\star)+\eta_\alpha ^{\ i}\partial_i\sK(z,z^\star)
\right]\right\}~,
\end{equation}
implying that $r_\alpha(z)$ describes the non-invariance of the K\"ahler
potential $\sK(z,z^\star)$ given by $\delta_\alpha \sK\equiv
\partial^i\sK\delta_\alpha z_i + \partial_i\sK\delta_\alpha z^i$. Note
that derivatives $\partial^i$ (and $\partial_i$) stand for derivatives
w.r.t.  $z_i$ (and $z^i\equiv z^\star$, respectively). Assuming that
the transformation of the conformon superfield $Y$ is given by
imaginary constants, then
\begin{equation}\label{transfoFI}
r_\alpha = i\frac{ g
\xi_\alpha}{3\Mpl^2}~~\mbox{and}~~\partial_i\xi_\alpha=0~;
\end{equation}
$g$ is the gauge coupling constant.  If $r_\alpha(z)\neq 0$, the
conformon superfield $Y$ transforms under a $U(1)$.  Only if
$r_\alpha(z)= 0$ the symmetry is preserved without corrections of the
superconformal U(1).  Equation (\ref{transfoFI}a) shows the
superconformal origin of FI terms via the gauge transformations of the
conformon field.  Equation (\ref{ralpha}) implies that the K\"ahler
potential is invariant~\cite{toine1}
\begin{equation}
\delta_\alpha\sK=0~.
\end{equation}
The function $\sW$ should be invariant under Yang-Mills
transformations,
\begin{equation}
\delta_\alpha\sW=0~,
\end{equation}
implying
\begin{equation}
\delta_\alpha W\equiv \eta_{\alpha i}\partial^i W=-3r_\alpha W~.
\label{noninv}
\end{equation}
Since the \emph{r.h.s.} of the above equation, Eq.~(\ref{noninv}), is
non-zero when FI terms are present, one concludes that the
superpotential $W$ cannot be gauge invariant, if the conformon
multiplet $Y$ transforms under gauge transformations. Assuming that
$r_\alpha$ is given by Eq.~(\ref{transfoFI}a), which corresponds to a
constant FI term in supergravity, one gets~\cite{toine1}
\begin{equation}\label{transfoW}
\delta_\alpha W=\eta_{\alpha i}\partial^i W = -i
\frac{g\xi_\alpha}{\Mpl^2}W~.
\end{equation}

To construct a formulation of supergravity with constant FI terms from
superconformal theory, one finds~\cite{toine1} that under U(1) gauge
transformations in the directions in which there are constant FI terms
$\xi_\alpha$, the superpotential $W$ must transform as in
Eq.~(\ref{transfoW}). As a consequence, we cannot keep any longer the
charge assignment described in the case of standard supergravity.

\subsubsection{Supergravity formulations from gauge fixing}

As it was explicitly shown in Ref.~\cite{toine1}, the gauge fixing
of the local dilatational invariance introduces the mass scale
$M_{\rm Pl}$, which we set $\Mpl^2=-\frac{1}{3}\sN$, and fixes the
value of the conformon superfield in terms of the K\"ahler potential
$\sK(z,z^\star)$, which depends only on the physical scalars $z$
and $z^\star$, namely
\begin{equation}
|Y|^2 = \Mpl^2\exp (\sK(z,z^\star)/3)~.
\end{equation}
The value $|Y|$ being fixed, we then fix the phase of $Y$.
The superconformal Lagrangian, Eq.~(\ref{lagrangSuperconf}),
is invariant under the redefinition
\begin{equation}
Y\rightarrow Y e^{\Lambda_Y(z)/3}~,
\end{equation}
for an arbitrary holomorphic function $\Lambda_Y(z)$, since this
redefinition can be absorbed in a redefinition of $z_i$. We can
check that the effect of this redefinition on $\sK$ and $W$ is
precisely the K\"ahler transformation introduced in
Eq.~(\ref{defkahltransf}).

Two choices have been proposed~\cite{toine1,SCFT} in order to fix
the phase of $Y$. Either
\begin{equation}
\sW = \sW ^\star~,
\label{choice1}
\end{equation}
leading to the standard formulation of SUGRA, described previously, or
alternatively
\begin{equation}
Y = Y^\star~.
\label{choice2}
\end{equation}
The first choice, Eq.~(\ref{choice1}), makes sense only for $\sW\neq
0$, while the second one, Eq.~(\ref{choice2}), is appropriate for
cases where $\sW$ can become zero.  Since in what follows we focus on
D-term inflation, where the superpotential vanishes during inflation
as well as at the end of the inflationary era, we adopt the second
choice of gauge fixing.  In this case, the form of the Lagrangian is
explicitly given in Refs.~\cite{toine1,SCFT}.  We give below several
quantities which are of interest for the purpose of our study.

To switch to dimensional K\"ahler potential and fields, we redefine
$z_i$ and $\sK$, as~\cite{toine1}
\begin{equation}
z_i=z_i^0+\frac{\Phi_i}{\Mpl}~~\mbox{and}\quad
\sK=\Mpl^{-2}K(\Phi_i,\Phi_i^\star,\Mpl^{-1})~;
\end{equation}
$K(\Phi_i,\Phi_i^\star,\Mpl^{-1})$ is regular at $M_{\rm
Pl}^{-1}=0$. We denote the scalar part of the superfield $\Phi_i$ by
$\phi_i$. From now on derivatives $\partial^i$ stand for derivatives
w.r.t. $\phi_i$; we thus use the fields $\phi_i$ and their complex
conjugates $\phi^i$ to indicate scalar fields.  The previous choice of
conformon transformation, Eq.~(\ref{transfoFI}), now reads
$\tilde{r}_\alpha (\phi)=ig\xi_\alpha/3$. The supergravity Lagrangian
depends on $W(\phi), K(\phi,\phi^\star)$ and $f_{ab}(\phi)$. Constant
FI terms can be introduced for some of the U(1) gauge groups.

We write below the bosonic and fermionic parts of the Lagrangian for
the scalar fields $\phi_i$, which are relevant for D-term inflation in
the context of supergravity formulated using superconformal
theory. The part of the bosonic sector which interests us,
is
\begin{equation}
e^{-1}\sL_{\rm bos}=-{g_i}^j(\partial_\mu\phi^i)(\partial^\mu
\phi_j)-V_F-V_D+\dots~,
\end{equation}
with
\begin{eqnarray}\label{potenFterms}
V_F&=& e^{K/\Mpl^2}\left[ (\sD^i W){(g^{-1})_i}^j(\sD_j W^*)
-3\frac{|W|^2}{\Mpl^2}\right]\nonumber\\
V_D&=&\frac{1}{2}\left[({\rm Re}f)^{-1}\right]^{\alpha\beta}
\sP_\alpha \sP_\beta~,
\end{eqnarray}
where
\begin{equation}
\sP_\alpha(\phi,\phi^\star,M_{\rm Pl}^{-1}) = i\left[ (\delta_\alpha
\phi_i) \partial^i K(\phi,\phi^\star,M_{\rm
Pl}^{-1})-3\tilde{r}_\alpha(\phi,M_{\rm Pl}^{-1}) \right]~,
\end{equation}
and
\begin{equation}
\sD^i W =\partial^i W+ \Mpl^{-2}(\partial^i K)W~.
\end{equation}

The fermionic mass terms for fermions $\chi_i$ of chiral
supermultiplets $\phi_i$ are given by the fermionic sector of the
Lagrangian, namely~\cite{toine1}
\begin{equation}\label{lagrangfermio}
e^{-1}\sL_{\rm ferm}=-{g_i}^j\left[ \bar{\chi}_j\Dbar
\chi^i+\bar{\chi}^i\Dbar \chi_j\right] -m^{ij}\bar{\chi}_i\chi_j -
m_{ij}\bar{\chi}^i\chi^j + e^{-1}\sL_{\rm mix}~,
\end{equation}
where $m\equiv e^{\sK/2}W$ and
\begin{eqnarray}
m^{ij}&\equiv& \sD^i\sD^j m=\left(\partial^i+\frac{1}{2}(\partial^i
\sK)\right)m^j -\Gamma^{ij}_k m^k\nonumber \\ 
m^i&\equiv& \sD^i m
=e^{\sK/2}\sD^iW=\partial^i m +\frac{1}{2}(\partial^i \sK)m \\
m_i&\equiv& \sD_i m^* =e^{\sK/2}\sD_i W^*=\partial_i
m^*+\frac{1}{2}(\partial_i \sK)m^*~,
\end{eqnarray}
and where the K\"ahler metric and K\"ahler connection are defined by
\begin{eqnarray}
{(g)_i}^j & \equiv& \partial_i\partial^j K\nonumber\\
\Gamma^{ij}_k & \equiv& (g^{-1})_k^l\partial^i g^j_l~.
\end{eqnarray}
In addition to the standard mass terms proportional to $m_{ij}$,
some mixing mass terms between the chiral fermions $\chi_i$ and
the gaugino or the gravitino are contained in $\sL_{\rm mix}$.
They can potentially contribute to the fermionic mass
matrix~\cite{toine1}:
\begin{eqnarray}\label{lagrangfermioMIX}
e^{-1}\sL_{\rm mix}=&&-2 m_{i\alpha}\bar{\chi}^i\lambda^\alpha -2
m^i_{\alpha} \bar{\chi}_i \lambda^\alpha \nonumber\\
&&+ \left[{g_j}^i
\bar{\psi}_{\mu L}(\partialbar
\phi^j)\gamma^{\mu}\chi_i+\bar{\psi}_R\cdot\gamma
v^1_L+h.c.\right]~,
\end{eqnarray}
where
\begin{eqnarray}
m_{i\alpha}&=&-i\left[\partial_i\sP_\alpha -
\frac{1}{4}(\mathrm{Re} f)^{-1\beta\gamma}\sP_\beta
f_{\gamma\alpha i}\right]\nonumber\\
v_L^1&=&\frac{1}{2}i\sP_\alpha\lambda_L^\alpha+m^i\chi_i~.
\end{eqnarray}
In the case of non vanishing $\sL_{\rm mix}$, the masses of are
obtained by diagonalising the fermionic mass matrix.

\subsubsection{Consequences for D-term inflation in SUGRA}\label{csqforSUGRA}
Let us investigate the consequences of a formulation of supergravity
constructed from superconformal theory for D-term inflation.  First,
the transformation of the SUGRA superpotential given by the
Eq.~(\ref{transfoW}) imposes a modification of the charge assignment
of Section~\ref{sec:stdSUGRAformul}. This charge assignment holds
for the global supersymmetric limit. Since the charge of the
superpotential is given by the sum of the charges of the fields $S$,
$\Phi_\pm$, we must impose~\cite{toine1}
\begin{equation}\label{chargassign}
q(S)= -\frac{\xi}{\Mpl^2}\rho_S, \quad q(\Phi_\pm)=\pm 1
-\frac{\xi}{\Mpl^2}\rho_\pm ~,  \quad \mathrm{where}\quad
\sum_{i=S,\pm} \rho_i =1~.
\end{equation}
Note that as a consequence, we can relate the FI amplitude to the
anomaly of the symmetry U(1)
\begin{equation}
\frac{\xi}{\Mpl^2}=-\ttr~Q=\sum_{i=S,\pm} q_i~,
\end{equation}
similarly to the generation of FI terms with anomalous U(1)
symmetry within weakly coupled string theories.

In Ref.~\cite{toine1} the authors suggest a way to solve the anomalous
U(1) symmetry. More precisely, the authors introduce new matter
superfields, which are not involved in inflation, with appropriate
charge assignments so that the anomaly gets
cancelled\footnote{Cancellation conditions for gauge anomalies within
$N=1$ four-dimensional supergravity with Fayet-Iliopoulos couplings
has been recently addressed in Refs.~\cite{freedman1,freedman2}.}. In
order to avoid an $\eta-$problem we set $\rho_S=0$. This modification
of the charge assignment induces a modification of the D-terms of the
scalar potential. Since the function $K$ is real and $W$ is
holomorphic, we can check that the expression of the F-terms,
Eq.~(\ref{potenFterms}), is identical to the standard formulation of
supergravity.

Concerning the radiative corrections, the modification of the
charges induces a modification of the mass terms for the scalar
components of the chiral superfields, since part of their mass
terms comes from the D-terms. The fermionic Lagrangian,
Eq.~(\ref{lagrangfermio}), is identical to the expression obtained
in the context of standard supergravity given
in~\cite{bailinlove,nilles}. Note that in the case of interest,
there are no additional contributions to the chiral fermion mass
terms from $\sL_{\rm mix}$ given in Eq.~(\ref{lagrangfermioMIX}).

The last consequence of the superconformal origin of supergravity
concerns the form of the superpotential. We first remark that the
transformation of the superpotential, Eq.~(\ref{transfoW}), combined
with the requirement that the inflaton field should remain uncharged
in order to avoid the $\eta$-problem, protect the form of the
superpotential
\begin{equation}\label{superpot}
W=\lambda S \Phi_+ \Phi_-~;
\end{equation}
all non-renormalisable terms of the form $S(\Phi_+
\Phi_-)^n/\Mpl^{2n-2}$ are forbidden. Secondly, if we assume in
addition that the superfields and the superpotential transform
under an R-symmetry as
\begin{eqnarray}
&&\Phi_+ \rightarrow e^{i\beta}\Phi_+~, \quad \Phi_- \rightarrow
e^{-i\beta}\Phi_-~, \nonumber\\ &&S \rightarrow e^{i\alpha} S~, \quad W
\rightarrow e^{i\alpha} W~,
\end{eqnarray}
then all terms proportional to $S^n$, with $n>1$, and all terms of
the form $f(S)g(\Phi_+)$ or $f(S)g(\Phi_-)$ are forbidden. We end
up with a D-term inflation that is precisely described by the
minimal superpotential, Eq.~(\ref{superpot}), even if in
supergravity non-renormalisable terms can be present. This
motivates our choice to consider in this work, the standard form
of superpotential.

\section{D-term inflation in minimal SUGRA}
The minimal supergravity description is based on the minimal
K\"ahler potential
\begin{equation}\label{Kmin}
K_{\rm min}=\sum_i |\Phi_i|^2=|\Phi_-|^2+|\Phi_+|^2+|S|^2~,
\end{equation}
and the minimal structure for $f_{ab}(\Phi_i)$, namely
$f_{ab}(\Phi_i)=\delta_{ab}$. The tree level scalar potential
reads\footnote{This formula differs slightly from the one in
Ref.~\cite{toine1}. In the last F-term there is a factor 3 instead of
6.}~\cite{toine1}
\begin{eqnarray}\label{DpotenSUGRAtotbis}
V_{\rm SUGRA}=&&
\lambda^2\exp\left({\frac{|\phi_-|^2+|\phi_+|^2+|S|^2}{M^2_{\rm
Pl}}}\right) \nonumber\\ && ~~~~~\times
\Biggl[|\phi_+\phi_-|^2\left(1+\frac{|S|^4}{M^4_{\rm
Pl}}\right)+|\phi_+S|^2 \left(1+\frac{|\phi_-|^4}{M^4_{\rm
Pl}}\right)\nonumber\nonumber\\
 && ~~~~~
+|\phi_-S|^2 \left(1+\frac{|\phi_+|^4}{M^4_{\rm
Pl}}\right) +3\frac{|\phi_-\phi_+S|^2}{M^2_{\rm Pl}}\Biggr]\nonumber \\
&&+ \frac{g^2}{2}\left(q_+|\phi_+|^2+q_-|\phi_-|^2+\xi\right)^2~.
\end{eqnarray}

The next step is the calculation of the masses of the components of
the superfields $\Phi_\pm$. The scalar mass squared can be read
directly from the scalar potential~\cite{toine1}, namely
\begin{equation}\label{scalmass}
m^2_\pm = \lambda^2 e^{|S|^2/\Mpl^2}|S|^2+g^2 q_\pm \xi~.
\end{equation}
Strictly speaking, one must use the charge assignment given by
Eq.~(\ref{chargassign}), for example,
\begin{equation}
q_\pm = \pm 1-\frac{1}{2}\frac{\xi}{\Mpl^2}~.
\end{equation}
Hereafter we assume $\xi/\Mpl^2\ll 1$ and therefore
the expression for the scalar masses, Eq.~(\ref{scalmass}),
can be approximated as
\begin{equation}\label{approxmass}
m^2_\pm \simeq \lambda^2 e^{|S|^2/\Mpl^2}|S|^2 \pm g^2 \xi~.
\end{equation}
In the inflationary valley, where $\phi_\pm=0$, the associated Dirac
fermions have a mass squared
\begin{equation}
m^2_{\rm f} = \lambda^2 e^{|S|^2/\Mpl^2}|S|^2~,
\end{equation}
unchanged compared to the case of standard minimal
supergravity~\cite{jcap2005,prl2005}.

Using the Coleman-Weinberg formula, we can then compute the
effective inflationary potential, taking into account the
tree-level and the one-loop radiative corrections due to the mass
splitting between components of the chiral superfields $\Phi_\pm$.
The effective potential reads~\cite{jcap2005,prl2005}
\begin{equation}\label{scalarpeff}
V^{\rm eff}(|S|)=\frac{g^2\xi^2}{2}\left\{
1+\frac{g^2}{16\pi^2}\left[2\ln\left(z\frac{g^2\xi}{\Lambda^2}\right)+f_V(z)
\right] \right\} ~,
\end{equation}
where
\begin{equation}
f_V(z) = (z+1)^2\ln\left( 1+\frac{1}{z}\right) + (z-1)^2\ln\left(
1-\frac{1}{z}\right)~,
\end{equation}
and
\begin{equation}
z\equiv \frac{\lambda^2}{g^2\xi} |S|^2
\exp\left(\frac{|S|^2}{M_{\rm Pl}^2}\right)~.
\end{equation}

\begin{figure}[hhh]
\begin{center}
\includegraphics[scale=.5]{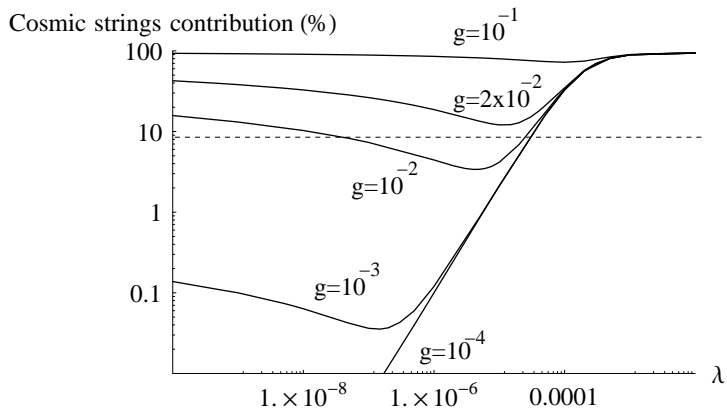}
\caption{In the framework of standard D-term inflation in minimal
SUGRA, cosmic string contribution to CMB quadrupole anisotropies, for
various values of the gauge coupling $g$, as a function of the
superpotential coupling constant $\lambda$.}\label{Fig:Dtermstandard}
\end{center}
\end{figure}

D-term inflation leads to cosmic strings formation at the end of the
inflationary era. Consistency between CMB measurements and theoretical
predictions constrain the parameters space. The constrains are
imposed on the couplings and mass scales. They are shown in
Fig.~\ref{Fig:Dtermstandard} and can be summarised
as~\cite{jcap2005,prl2005}
\begin{equation}\label{constrDmin}
g\lsim 2\times 10^{-2}~ ~ \mbox{and}~ ~
\lambda\lsim 3\times 10^{-5}~.
\end{equation}
The above constraints, Eq.~(\ref{constrDmin}), can be
expressed~\cite{jcap2005,prl2005} as a single constraint on the
Fayet-Iliopoulos term $\xi$, namely,
\begin{equation}
\sqrt\xi \lsim 2\times 10^{15}~{\rm GeV}~.
\end{equation}
We can therefore see that our assumption that
$\xi/\Mpl^2\sim 10^{-6} \ll 1$ in Eq.~(\ref{approxmass}) is
justified~\cite{msadp,msspringer,procRocher}, which implies that for
the purpose of our study and for minimal D-term inflation, we can
neglect the corrections introduced by the superconformal origin of
supergravity. Moreover, since the tree level potential is given by
$V_0\propto \xi^2$, the limit $\xi/\Mpl^2 \ll 1$ should hold since
otherwise the energy density would become transplankian and the
quantum gravity corrections, which have been so far
neglected~\cite{jcap2005}, would become important~\cite{lindebook}.\\

In what follows we address whether the restrictions found in
the framework of minimal SUGRA, are still qualitatively valid for
a non-minimal SUGRA theory. We will thus study D-term hybrid
inflation with a superpotential $W$ defined by
Eq.~(\ref{superpoteninflaD}) and different choices for the form of
the K\"ahler potential.

\section{Inflation with shift symmetry}
We first study hybrid inflation in the context of supergravity,
with a K\"ahler potential obeying a shift symmetry. This symmetry
can be used for model building in the framework of supergravity,
while it is also motivated from string theory~\cite{linde05}. More
precisely, the shift symmetry,
\begin{equation}
S \rightarrow S + i C~,
\end{equation}
with $C$ a real constant, has been used in order to obtain flat
potentials from F-terms in SUGRA, so that the $\eta$-problem is
cured, or to make compatible chaotic inflation, which requires an
inflaton field larger than the Planck scale, and
supergravity~\cite{shiftK}.

Thus, we consider the following form for the K\"ahler potential
\begin{equation}\label{K3}
K_1=\frac{1}{2} (S+\bar{S})^2+|\phi_+|^2+|\phi_-|^2~,
\end{equation}
and minimal structure for the kinetic function. We note that
compared to the minimal K\"ahler potential, Eq.~(\ref{Kmin}), $K_1$
has two additional terms,, $S^2/2$ and $\bar{S}^2/2$, which do not
affect the kinetic terms.

Using the standard expression for the scalar potential,
Eq.~(\ref{formPotenScal}), we obtain,
\begin{equation}
V=V_F+V_D~,
\end{equation}
where the D-terms are not modified as compared to the minimal
SUGRA case. Thus, assuming $\xi/\Mpl^2 \ll 1$,
\begin{equation}
V_D \simeq \frac{g^2}{2}\left(|\phi_+|^2-|\phi_-|^2+\xi\right)^2
~,
\end{equation}
and the F-terms are given by
\begin{eqnarray}
V_{\rm F}=&&
\lambda^2\exp\left({\frac{|\phi_-|^2+|\phi_+|^2}{M^2_{\rm
Pl}}}\right)\exp\left[{\frac{(S+\bar{S})^2}{2M^2_{\rm Pl}}}\right]
\nonumber\\ & & ~~~\times
\Biggl[|\phi_+\phi_-|^2\left(1+\frac{S^2+\bar{S}^2}{M^2_{\rm
Pl}}+\frac{|S|^2|S+\bar{S}|^2}{M^4_{\rm Pl}}\right)+|\phi_+S|^2
\left(1+\frac{|\phi_-|^4}{M^4_{\rm Pl}}\right) \nonumber\\ &&
~~~+|\phi_-S|^2 \left(1+\frac{|\phi_+|^4}{M^4_{\rm Pl}}\right)
+3\frac{|\phi_-\phi_+S|^2}{M^2_{\rm Pl}}\Biggr] ~.
\end{eqnarray}
As in the case of D-term inflation studied within minimal
supergravity, the potential has a global minimum
\begin{equation}
V=0 \quad \mathrm{for}\quad \langle S\rangle=0~,\;
\langle\Phi_+\rangle=0~, \; \langle\Phi_-\rangle=\sqrt{\xi}~.
\end{equation}
There is also a local minimum for large values of $S$,
\begin{equation}
V=\frac{g^2\xi^2}{2} \quad \mathrm{for}\quad \langle S\rangle\gg
S_c~,\; \langle\Phi_\pm\rangle=0~.
\end{equation}

\begin{figure}[hhh]
\begin{center}
\includegraphics[scale=.5]{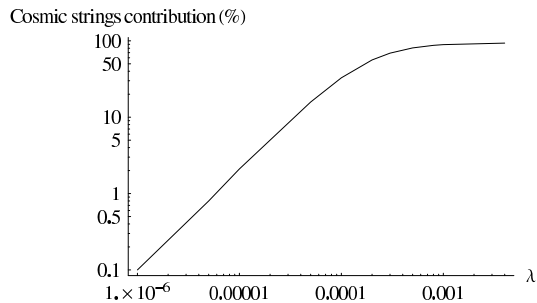}
\caption{Cosmic string contribution to the CMB temperature
anisotropies as a function of the superpotential coupling constant
$\lambda$, in the case of the D-term inflation model in the framework
of global SUSY.  Figure taken from Ref.~\cite{jcap2005}.}
\label{Dtermsusy}
\end{center}
\end{figure}

Comparing with the scalar potential obtained in minimal SUGRA, we
can identify two differences. The first one is that there are
several new terms proportional to $|\Phi_+\Phi_-|^2$. These terms
will not affect the effective mass terms of the $\Phi_\pm$ fields,
thus they will not affect the inflationary potential, even when
one-loop radiative corrections have been taken into account. The
second one is that the exponential factor $e^{|S|^2}$ in minimal
SUGRA, has been here replaced by $e^{(S+\bar{S})^2/2}$.  Writing
$S=\eta+i\phi_0$, we get $e^{(S+\bar{S})^2/2}=e^{\eta^2}$.  If we
identify the inflaton field with the real part of $S$, then we
obtain exactly the same potential as in the minimal case. However,
with the present choice for the K\"ahler potential, $K_1$, it would
be more logical to identify the inflaton field with the imaginary
part of $S$, namely $\phi_0$. The reason being that with this choice the
exponential term is constant during inflation, thus it cannot
spoil the slow roll conditions. Then we can check that the
inflationary potential we get is identical to the usual D-term
inflation within the global SUSY framework. This model has been
studied in Ref.~\cite{jcap2005}. The result obtained in
Ref.~\cite{jcap2005} is reminded in Fig.~\ref{Dtermsusy}.
Clearly, in such a model the cosmic string contribution to the
CMB anisotropies is dominant, in contradiction with the CMB
measurements, unless the superpotential coupling is constrained to be
\begin{equation}
\lambda\lsim 3\times 10^{-5}~.
\end{equation}
Concluding, we state that simply imposing a shift symmetry to the
K\"ahler potential is not enough to {\sl escape} the {\sl cosmic
string problem} of D-term inflation.

\section{D-term inflation with higher order terms}\label{secpotenseto}
In this section, we consider another choice of K\"ahler potential that
contains non renormalisable terms. More precisely, we consider the
following form
\begin{equation}
K_2=|S|^2+|\Phi_+|^2+|\Phi_-|^2+f_+\bigg(\frac{|S|^2}{M_{\rm
Pl}^2}\bigg)|\Phi_+|^2+f_-\bigg(\frac{|S|^2}{M_{\rm Pl}^2}\bigg)
|\Phi_-|^2+b\frac{|S|^4}{\Mpl^2}~,
\label{gen}
\end{equation}
where $f_\pm$ are arbitrary functions of $(|S|^2/M_{\rm Pl}^2)$.
Contrary to the previous section, this new form should modify the
kinetic terms. The motivation for this choice is that it contains
the next-to-minimal K\"ahler potential, with all terms up to order
$\Mpl^{-2}$. We consider in a first place a reduced version of
this potential and then we proceed with the full expression. We
remind to the reader that our primary goal is not to build a new
model for D-term inflation in the context of supergravity, but to
study the robustness of our prediction that there is a dominant
contribution of cosmic strings to the CMB temperature
anisotropies, unless we fine tune the superpotential coupling
constant $\lambda$.

\subsection{Simplified case : $b=0$}
We first consider the case of the previous K\"ahler potential,
Eq.~(\ref{gen}), setting $b=0$, implying
\begin{equation}
K_3=|S|^2+|\Phi_+|^2+|\Phi_-|^2+f_+\bigg(\frac{|S|^2}{M_{\rm
Pl}^2}\bigg)|\Phi_+|^2+f_-\bigg(\frac{|S|^2}{M_{\rm Pl}^2}\bigg)
|\Phi_-|^2~, \label{kahlerseto1}
\end{equation}
where $f_\pm$ are arbitrary functions of $(|S|^2/M_{\rm Pl}^2)$,
while the superpotential is given in Eq.~(\ref{superpoteninflaD}).
One can argue that the K\"ahler potential given in
Eq.~(\ref{kahlerseto1}) is quite general, since during inflation
the Higgs fields are small. However, one can also criticise this
choice for the following reason: even though $|\phi_\pm|^4$-terms
are indeed negligible as compared to the $|\phi_\pm|^2$-ones (the
$|\phi_\pm|$-terms are small during the inflationary era) there is
no reason for neglecting $|S|^4/M_{\rm Pl}^4$-terms. They will be
taken into account in next section.

In this model, the scalar potential can be calculated using
Eq.~(\ref{formPotenScal}) and reads
\begin{equation}\label{DpotenSUGRAnonmin}
V(|S|)=V_F+V_D~,
\end{equation}
where the F-part is\footnote{We disagree with the expression given in
Ref.~\cite{seto}. We believe that in Ref.~\cite{seto} the authors made
some unjustified assumptions under which our expression for $V_F$,
given in Eq.~(\ref{ourvf}), can be reduced to the expression given in
Ref.~\cite{seto}. Even though this disagreement has no implications
for the rest of our study, we would like to bring to the attention of
the reader that the correct expression for the F-contribution to the
scalar potential, $V_F$, is indeed given in Eq.~(\ref{ourvf}).}
\begin{eqnarray}
V_F=&&\lambda^2\frac{e^K}{{\rm det} {K_i}^j}
\bigg[\bigg\{(1+f_+)(1+f_-)\nonumber\\
&&~~~~~~~~~~~\times\left[1+\frac{|S|^2}{\Mpl^2}
\left(1+f'_+\frac{|\phi_+|^2}{\Mpl^2}+f'_-\frac{|\phi_-|^2}
{\Mpl^2}\right)\right]^2\bigg\}|\phi_+\phi_-|^2\nonumber\\
&&+\bigg\{\left[(1+f_+)(1+\Delta)-{f'_+}^2
\frac{|S\phi_+|^2}{\Mpl^4}\right]\left[1+\frac{|\phi_-|^2}
{\Mpl^2}(1+f_-)\right]^2\bigg\}|\phi_+S|^2\nonumber\\
&&+\bigg\{\left[(1+f_-)(1+\Delta)-{f'_-}^2
\frac{|S\phi_-|^2}{\Mpl^4}\right]
\left[1+\frac{|\phi_+|^2}{\Mpl^2}(1+f_+)\right]^2\bigg\}|\phi_-S|^2\nonumber\\
&&-2\bigg\{\left[1+\frac{|S|^2}{\Mpl^2}
\left(1+f'_+\frac{|\phi_+|^2}{\Mpl^2}+f'_-\frac{|\phi_-|^2}{\Mpl^2}\right)\right]
\nonumber\\
&&~~~~~~~~~~\times\left[(1+f_+)(1+f_-)\right]'\bigg\}\frac{|\phi_+\phi_-S|^2}{\Mpl^2}\nonumber\\
&&+2\bigg\{f'_+f'_-|S|^2\left[1+\frac{|\phi_+|^2}{\Mpl^2}(1+f_+)\right]
\left[1+\frac{|\phi_-|^2}{\Mpl^2}(1+f_-)\right]\bigg\}\frac{|\phi_+\phi_-S|^2}{\Mpl^2}\bigg]
\nonumber\\
&&-3\lambda^2 e^K \frac{|S\phi_+\phi_-|^2}{\Mpl^2}~.
\label{ourvf}
\end{eqnarray}
and the D-part is
\begin{equation}\label{VDnonmin}
V_D=\frac{g^2}{2}\bigg[q_+(1+f_+)|\phi_+|^2+q_-(1+f_-)|\phi_-|^2
+\xi\bigg]^2~.
\end{equation}
We have used the following notations
\begin{eqnarray}
f_\pm &\equiv &f_\pm (x)\bigg|_{x=|S|^2/M_{\rm Pl}^2}~,\nonumber\\
f'_\pm &\equiv& \frac{d f_\pm (x)}{dx}\bigg|_{x=|S|^2/M_{\rm Pl}^2}~,
~ f''_\pm \equiv \frac{d^2 f_\pm (x)}{dx^2}\bigg|_{x=|S|^2/M_{\rm
Pl}^2}~,\nonumber\\
\Delta & \equiv &\left(f'_+ + f''_+ \frac{|S|^2}{\Mpl^2}\right)
\frac{|\phi_+|^2}{\Mpl^2}+ \left(f'_- + f''_-
\frac{|S|^2}{\Mpl^2}\right) \frac{|\phi_-|^2}{\Mpl^2}
\end{eqnarray}
and
\begin{eqnarray}
{\rm det} K_i^{~ j}= &&~ (1+\Delta)(1+f_+)(1+f_-)\nonumber\\
&&-(1+f_+){f'_-}^2\frac{|S{\phi_-}|^2}{\Mpl^4}
-(1+f_-){f'_+}^2\frac{|S{\phi_+}|^2}{\Mpl^4}~.
\end{eqnarray}
As in minimal supergravity, the tree level inflationary potential
is constant and equal to
\begin{equation}
V_0=\frac{1}{2}\,g^2 \xi^2~, \label{treelevelsc}
\end{equation}
while supersymmetry is broken. The spontaneous symmetry breaking of
supersymmetry in the inflationary valley introduces a splitting in the
masses of the components of the chiral superfields $\Phi_\pm$. As a
result, one obtains two scalars with squared masses
\begin{equation}
m^2_\pm=\lambda^2|S|^2\exp\bigg(\frac{|S|^2}{M_{\rm Pl}^2}\bigg)
\frac{1}{(1+f_+)(1+f_-)}\pm g^2q_\pm\xi
\end{equation}
and a Dirac fermion with squared mass
\begin{equation}
m^2_{\rm fermion}= \lambda^2 |S|^2\exp\bigg(\frac{|S|^2}{M_{\rm
Pl}^2}\bigg) \frac{1}{(1+f_+)(1+f_-)}~.
\end{equation}
The one-loop radiative corrections to the inflationary potential
can be  calculated using the Coleman-Weinberg formula. Taking
also into account the tree level contribution, the effective
scalar potential for the considered D-term inflationary model
reads~\cite{seto}
\begin{equation}
V_{\rm eff}(|S|)=\frac{g^2\xi^2}{2}\left\{
1+\frac{g^2}{16\pi^2}\left[ 2\ln \left(
z\frac{g^2\xi}{\Lambda^2}\right)+f_V(z) \right] \right\} ~,
\end{equation}
with
\begin{equation}\label{deffV}
f_V(z) = (z+1)^2\ln\left( 1+\frac{1}{z}\right) + (z-1)^2\ln\left(
1-\frac{1}{z}\right)~
\end{equation}
and
\begin{equation}
z\equiv \frac{\lambda^2|S|^2}{g^2\xi}\exp\bigg(\frac{|S|^2}{M_{\rm
Pl}^2}\bigg) \frac{1}{(1+f_+)(1+f_-)}~.
\end{equation}
Strictly speaking the above expression holds for the limit where
$\xi/M_{\rm Pl}^2\ll 1$, which is indeed the case as one can
confirm in the end of this section.

Note that the expression for the effective scalar potential is
identical to that of minimal supergravity, except for the
expression of $z$. The first derivative of the potential is equal
to
\begin{equation}
V'_{\rm eff}(|S|)\equiv\frac{dV_{\rm eff}}{d|S|} = \frac{g^4
\xi^2}{16\pi^2}\, z f_z(|S|) f_{V'}(z) ~, \label{V'eff}
\end{equation}
where
\begin{equation}\label{deffV'}
f_{V'}(z)\equiv (z+1)\ln\left( 1+\frac{1}{z}\right) +
(z-1)\ln\left( 1-\frac{1}{z}\right) ~,
\end{equation}
\begin{equation}
f_{z}(|S|)\equiv 2|S|\left[ \frac{1}{M_{\rm Pl}^2}+\frac{1}{|S|^2}
- \frac{f'_{+}}{(1+f_+)} - \frac{f'_{-}}{(1+f_-)}\right] ~;
\end{equation}
$f'_{\pm}$ denote the first derivative of $f_\pm$ with respect to
$|S|^2$.  A choice which we will later consider for $f_{\pm}$ is
\begin{equation}
f_\pm\bigg(\frac{|S|^2}{M_{\rm Pl}^2}\bigg)=c_\pm
\frac{|S|^2}{M_{\rm Pl}^2}~, \quad \mathrm{therefore}\quad
f'_\pm=c_\pm\frac{1}{M_{\rm Pl}^2}~.
\end{equation}
In the large $|S|$-limit the effective potential and its first
derivative with respect to $|S|$ reduce to
\begin{equation}
V_{\rm eff}(|S|)\simeq\frac{g^4\xi^2}{16\pi^2}\left[ \ln \left(
\frac{\lambda^2|S|^2}{(1+f_+)(1+f_-)\Lambda^2}\right)+\frac{|S|^2}{M_{\rm
Pl}^2} \right]
\label{limitz1}
\end{equation}
and
\begin{equation}
 \frac{dV_{\rm
eff}}{d|S|}\equiv V'_{\rm
eff}(|S|)\simeq\frac{g^4\xi^2}{16\pi^2}f_z(|S|)~, \label{limitz2}
\end{equation}
respectively.

At this point, we would like to note that D-term inflation can be
realised with the last 60 e-folds being very close to the critical
value $z_{\rm end}= 1$, implying that the above reduced formulae,
Eqs.~(\ref{limitz1}), (\ref{limitz2}), \emph{cannot} be used to compute the
predictions of the model regarding the CMB temperature anisotropies.
Therefore, we disagree with the approach of Ref.~\cite{seto},
where the authors have used the reduced formulae for the
effective potential and its first derivative, to compute the
inflaton contribution to the CMB temperature anisotropies.

\begin{figure}[hhh]
\begin{center}
\includegraphics[scale=.5]{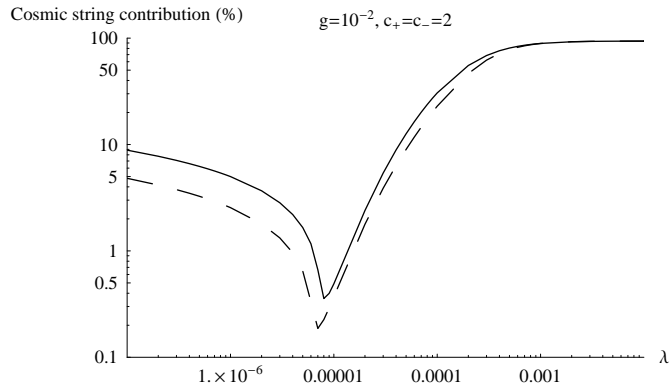}
\caption{Cosmic string contribution to the CMB quadrupole
anisotropies as a function of the superpotential coupling constant
$\lambda$ with partial (dashed line) or full (continuous line) scalar
potential.}\label{Fig:error}
\end{center}
\end{figure}

To illustrate the above remark, we represent in Fig.~\ref{Fig:error}
the error made by this assumption.  The cosmic string contribution is
drawn as a function of the superpotential coupling constant $\lambda$,
considering either the partial or the full scalar potential.  One can
realise that the error on the cosmic string contribution can reach
$50\%$, leading to a considerable error in the induced constraints on
the parameters space.

The number of e-foldings of inflation between the initial and
final value of the inflaton field is
\begin{equation}\label{efoldings}
N_{\rm Q}\equiv \ln \left( \frac{a_{\rm end}}{a_Q}\right) =
\frac{8\pi^2}{g^2 M_{\rm Pl}^2}\int_1^{z_Q} \frac{\dd z}{z^2
f^2_z[S(z)] f_{V'}(z)}~,
\end{equation}
where we note that the index $_{\rm Q}$ denotes the scale
responsible for the quadrupole anisotropy in the CMB.  For the
integral appearing in the above expression for $N_{\rm Q}$ to be
correctly defined, one must assume that the function
$$z\equiv \frac{\lambda^2}{g^2\xi} |S|^2 \exp\left(\frac{|S|^2}{M_{\rm
Pl}^2}\right)$$ can be inverted. Since it is a monotonic and
increasing function, one can verify, at least numerically, that
the function $z(|S|)$ can be inverted. We denote the inverted
function by $S(z)$.

The above expression for $N_{\rm Q}$, Eq.~(\ref{efoldings}), will
enable us to write a relation between $\xi$ and $z_{\rm Q}$.
Setting $N_{\rm Q}=60$, we can fix the value of the inflaton field
$S_{\rm Q}$, responsible for the quadrupole anisotropy of the CMB.

\subsubsection{Normalisation to COBE}
D-term inflation leads generically to cosmic strings formation,
which also contribute to the CMB temperature anisotropies.
The total contribution, from the inflaton field as well as from
the cosmic strings, has to be normalised to the value measured by
the COBE-DMR experiment. This normalisation will fix the mass
scale $\sqrt{\xi}$ as a function of the superpotential coupling
$\lambda$ and the gauge coupling $g$, which are considered here as
parameters. Thus, we have to calculate both, the cosmic strings
contribution and the inflaton field one, and normalise their
\emph{sum} to COBE-DMR. Here we emphasise that different
approaches can be found in the literature where the authors do NOT
normalise the sum of the two contributions to the data. Clearly
not to normalise the sum to the data is a simplification, which
may lead to an important error for the result. Normalising only the
inflationary contribution to the quadrupole absolute value can be
considered as an assumption where the cosmic string contribution
is neglected. This assumption could be {\sl allowed} only if one
wants to argue that the cosmic strings are sub-dominant. However,
since we already know that cosmic strings are indeed sub-dominant
and our aim is to constrain the parameters space and calculate
exactly how sub-dominant the strings are, this simplification in
the normalisation implies important errors in the calculated
cosmic string contribution.

From the scalar potential given in Eq.~(\ref{DpotenSUGRAnonmin}), one
can see that the spontaneous symmetry breaking  is generated when
$\phi_\pm$ takes a vacuum expectation value  $\sqrt{\xi}$. Thus,
the quadrupole contribution to the CMB temperature anisotropies from
the cosmic strings formed at the end of hybrid inflation, which is
\begin{equation}
\left(\frac{\delta T}{T}\right)_{\mathrm{Q-CS}} \simeq (9-10) G\mu
~~{\rm with}~~\mu=2\pi\langle {\cal {X}}\rangle^2~,
\end{equation}
where ${\cal {X}}$ is the vacuum expectation value of the Higgs
field responsible for the formation of cosmic strings,  is
approximately equal to
\begin{equation}
\left(\frac{\delta T}{T}\right)_{\mathrm{Q-CS}} \simeq
\frac{9}{4}\xi ~.
\end{equation}

For the contribution from the inflaton field, we evaluate the
Sachs-Wolfe term splitted into the scalar and tensor parts, using
\begin{equation}
\left(\frac{\delta T}{T}\right)_{\mathrm{Q-scal}} \simeq
\frac{1}{4\sqrt{45} \pi} \frac{V^{3/2}(S_{\rm Q})}{M_{\rm
Pl}^3V'(S_{\rm Q})} \label{inflsc}
\end{equation}
and
\begin{equation}
\left(\frac{\delta T}{T}\right)_{\mathrm{Q-tens}} \simeq
\frac{(0.77)}{(8\pi)} \frac{V^{1/2}(S_{\rm Q})}{M_{\rm Pl}^2}~.
\label{inflsctens}
\end{equation}
From Eqs.~(\ref{treelevelsc}), (\ref{V'eff}), (\ref{inflsc}) and
(\ref{inflsctens}) we get
\begin{equation}\label{dTsurT}
\left(\frac{\delta T}{T}\right)_{\mathrm{Q-scal}} \simeq
\frac{\sqrt{2}\pi}{\sqrt{45}}\frac{\xi}{g}\, \frac{1}{M_{\rm
Pl}^3}\, z_{\rm Q}^{-1}f_{V'}^{-1}(z_{\rm Q})f_z^{-1}(S_{\rm Q})
\end{equation}
and
\begin{equation}
\left(\frac{\delta T}{T}\right)_{\mathrm{Q-tens}} \simeq
\frac{0.77}{8\sqrt{2}\pi}\frac{1}{M_{\rm Pl}^2} g \xi~.
\end{equation}
We note that the ratio of the tensor part of the
inflaton field contribution to the cosmic strings one is constant,
and that the tensor part contribution can be neglected. The tensor
over scalar ratio, $r_{\rm infl}$, is less straightforward and it
is given by
\begin{equation}
r_{\rm infl} = \frac{0.77\sqrt{45}}{16\pi^2}\, g^2\, z_{\rm Q} \,
M_{\rm Pl}\, f_{V'}(z_{\rm Q}) f_z(S_{\rm Q})~.
\end{equation}

\begin{figure}[hhh]
\begin{center}
\includegraphics[scale=.5]{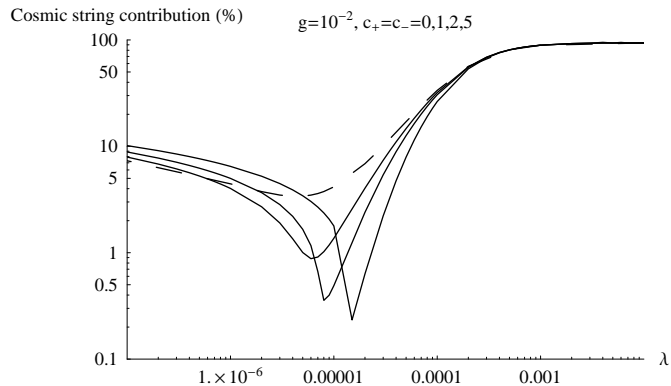}
\caption{The dashed line corresponds to the minimal SUGRA case.
The other lines correspond to various values of the parameter
$c=c_\pm=1,2,5$ from the top to the bottom, respectively.
This is for $g=10^{-2}$.}\label{CSdelambda}
\end{center}
\end{figure}

We then proceed with the cosmic string contribution to the
quadruple CMB temperature anisotropy. This contribution in
computed as a function of the superpotential coupling $\lambda$,
for various values of $g$ and $c=c_\pm$, which are considered as
parameters. The results are plotted in Fig.~\ref{CSdelambda}.

\begin{figure}[hhh]
\begin{center}
\includegraphics[scale=.5]{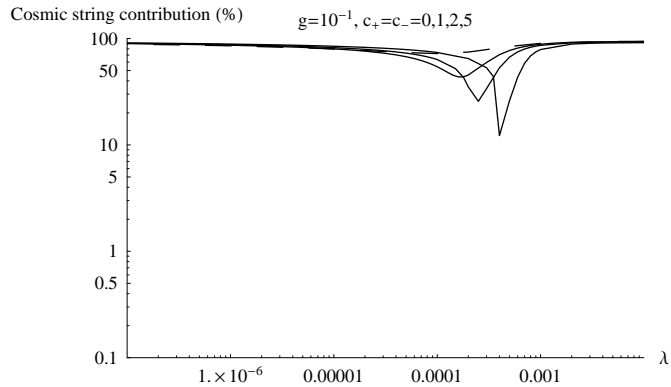}
\caption{The dashed line corresponds to the minimal SUGRA case.
The other lines correspond to various values of the parameter
$c=c_\pm=1,2,5$ from the top to the bottom, respectively.
This is for $g=10^{-1}$.} \label{CSdelambdag1}
\end{center}
\end{figure}

To show the dependence of the cosmic string contribution on the
gauge coupling $g$, we draw in Fig.~\ref{CSdelambdag1} the cosmic
string contribution for $g=10^{-1}$ and $c=c_\pm=0, 1, 2, 5$.
Clearly, this case ($g=10^{-1}$) is excluded since the cosmic
string contribution is above the allowed one. We note that we do
not take $c$ higher than 5, since positivity condition $V'(S)>0$
requires $c<3+2\sqrt{2}$~\cite{seto}.

We quantify below the constraints on the parameters space imposed from
the three-year Wilkinson Microwave Anisotropy Probe (WMAP)~\cite{wmap}
measurements. We want to set precise constraints on the free
parameters since this can be of importance for concluding of whether
D-term inflation remains in agreement with CMB data once we also
include constraints imposed from the allowed value of the spectral
index~\cite{inprep}.

For $g\gsim 10^{-1}$, there is no parameters space in agreement with
measurements. In the range $g\in [2\times 10^{-2},10^{-1}]$, the
parameters space is extremely small, around $\lambda\sim 2\times
10^{-5}$; the presence of $c$ enlarges slightly the allowed
window. For $g=10^{-2}$, the $9\%$ upper limit on the allowed cosmic
string contribution to the CMB imposes, at $95\%$ of confidence
level, the following constraints:
\begin{equation}\label{contraintelambdanonmin}
\left.\begin{array}{ll}
3\times 10^{-8} \\
5\times 10^{-8} \\
9\times 10^{-8} \\
2\times 10^{-7}\\
\end{array}
\right\} \leq \lambda \leq \left\{
\begin{array}{ll}
2.5\times 10^{-5} & \quad \mathrm{for}\quad c=0 \\
3.5\times 10^{-5} & \quad \mathrm{for}\quad c=1 \\
4.0\times 10^{-5} & \quad \mathrm{for}\quad c=2 \\
5.3\times 10^{-5} & \quad \mathrm{for}\quad c=5 \\
\end{array}
\right.
\end{equation}
or, equivalently,
\begin{equation}
\sqrt{\xi}\leq 2.2\times 10^{15}
\;\mathrm{GeV}  \Longleftrightarrow G\mu \leq 8.4\times 10^{-7}~.
\end{equation}
Clearly the new degree of freedom, namely $c=c_\pm$, allows a slightly
higher upper bound on the coupling, which is however at best higher by
only a factor of 2, thus concluding that fine tuning is still
required. All constraints are equivalent to a single constraint on
$\sqrt\xi$, or $G\mu$, as already stated in
Refs.~\cite{jcap2005,prl2005}.  Therefore, there is a bijection
between the cosmic string contribution and the mass scale of
inflation, which nevertheless does not mean that one can normalise
only the inflaton contribution to the CMB data. Clearly, the error
made by this assumption is big when cosmic strings have an important
weight, while it is small if the cosmic string contribution is small,
of the order the one  found here. With the current upper limit of
$9\%$ on the allowed cosmic string contribution to the temperature
anisotropies, the relative error made on the string contribution, by
not normalising the sum, is of the order of $10\%$.

\begin{figure}[hhh]
\begin{center}
\includegraphics[scale=.5]{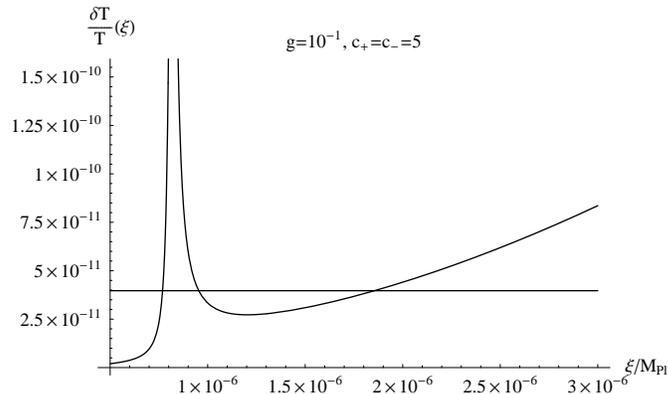}
\caption{Three solutions for the function ${\cal F}(\xi)$. In
terms of the cosmic string contributions these solutions
correspond to $7.5 \%, 11.6\%$ and $43.9\%$ from left to right,
equivalently.} \label{multval}
\end{center}
\end{figure}

As one can see from Figs.~\ref{CSdelambda}, \ref{CSdelambdag1} the
curve which showing the cosmic string contribution becomes singular
at its minimum, for large values of $c_\pm$. The reason for this
behaviour is that the function ${\cal F}(\xi)$, denoting the sum of
the cosmic strings and inflaton contributions as a function of $\xi$
becomes non-bijective for certain values of $\lambda$. This implies
that there are more than one solutions for the function ${\cal
F}(\xi)$ normalised to the COBE data, for a tiny window of the
parameter $\lambda$ around the value $10^{-5}$.  We illustrate this in
Fig.~\ref{multval}.
We have checked that this {\sl degeneracy} does not influence the
validity of our results since the different solutions for the
normalisation to COBE are of the same order of magnitude and this
behaviour is observed for a tiny parameters space.

As a final remark, we would like to note that there is also
another point where we disagree with Ref.~\cite{seto}.  In their
model specified by the choice of the K\"ahler potential and with the
simplifications made for the scalar potential, the authors
obtain~\cite{seto} a value for $\sqrt\xi$, required to generate
the appropriate magnitude of density perturbations. The authors
argue that this value is consistent with the upper limit imposed
on $\xi$ so that the cosmic string contribution is within the
allowed window, following the results of Ref.~\cite{endoetal}.
However, the analysis of Ref.~\cite{endoetal} was done employing
the curvaton mechanism; a contribution which the authors of
Ref.~\cite{seto} have not considered at all.

\subsection{General case}
We proceed with the general case, namely we include all terms of the
next to leading order
\begin{equation}
K_2=|S|^2+|\Phi_+|^2+|\Phi_-|^2+f_+\bigg(\frac{|S|^2}{M_{\rm
Pl}^2}\bigg)|\Phi_+|^2+f_-\bigg(\frac{|S|^2}{M_{\rm Pl}^2}\bigg)
|\Phi_-|^2+b\frac{|S|^4}{\Mpl^2}~,
\end{equation}
where the function $f_\pm$ is just
\begin{equation}
\label{fpm}
f_\pm\bigg(\frac{|S|^2}{M_{\rm Pl}^2}\bigg)=c_\pm \frac{|S|^2}{M_{\rm
Pl}^2}~.
\end{equation}

The motivation for such a choice is that beyond the minimal
supergravity part of the K\"ahler potential, the leading order is
$\Mpl^{-2}$, implying that one should consider all
corrections up to this order. During inflation, the charged fields
$\Phi_\pm$ vanish, meaning that terms of the form
$|\Phi_\pm|^4/\Mpl^2$ or $|\Phi_+|^2|\Phi_-|^2/\Mpl^2$ can be
neglected compared to $|\Phi_\pm|^2$-terms, which are in $K_{\rm
min}= |S|^2+|\Phi_+|^2+|\Phi_-|^2$.  Thus, the next-to-minimal
D-term inflationary model should be constructed with K\"ahler
potential $K_2$. We again consider the superpotential
\begin{equation}
W=\lambda S \Phi_+ \Phi_-~,
\end{equation}
involving the superfields $S$ and $\Phi_\pm$ with charges $0$ and
$q_\pm$, respectively, under the $U(1)_\xi$ symmetry of the
Lagrangian. The charges $q_\pm$ must satisfy the constraint induced by
the superconformal origin of the FI term, as discussed in
Section~\ref{csqforSUGRA}.  The D-term part of the scalar potential is
thus unchanged as compared the previous section and given by
Eq.~(\ref{VDnonmin}).

We first compute the inverse K\"ahler metric $g^{-1}$, defined by
${g^i}_j {(g^{-1})^j}_k={\delta^i}_k$: 
\begin{equation}\label{Kmetricinvers}
{(g^{-1})^j}_i=\frac{1}{g}\times {\cal M}~,
\end{equation}
where ${\cal M}$ is equal to
\begin{equation}
{\hskip-2.6truecm}
\left(
\begin{array}{ccc}
(1+f_+)(1+f_-) &
-c_+\frac{S^*\phi_+}{\Mpl^2}(1+f_-) &
-c_-\frac{S^*\phi_-}{\Mpl^2}(1+f_+)\\
-c_+\frac{S\phi_+^*}{\Mpl^2}(1+f_-) & \left(1+\Delta\right)(1+f_-)
-c_-^2\frac{|S|^2|\phi_-|^2}{\Mpl^4}&
c_+c_-\frac{|S|^2}{\Mpl^2}\frac{\phi_+^*\phi_-}{\Mpl^2} \\
-c_-\frac{S\phi_-^*}{\Mpl^2}(1+f_+) &
c_+c_-\frac{|S|^2}{\Mpl^2}\frac{\phi_+\phi_-^*}{\Mpl^2} &
\left(1+\Delta\right)(1+f_+) -c_+^2\frac{|S|^2|\phi_+|^2}{\Mpl^4}
\end{array}\right)
\end{equation}
and we have used Eq.~(\ref{fpm}) and the definitions
\begin{equation}
\Delta \equiv
c_+\frac{|\phi_+|^2}{\Mpl^2}
+c_-\frac{|\phi_-|^2}{\Mpl^2}+2b\frac{|S|^2}{\Mpl^2}~;\nonumber
\end{equation}
\begin{eqnarray}
g\equiv \det
(g)_i^j=&&~~\left(1+\Delta\right)(1+f_+)(1+f_-)
\nonumber\\&&-(1+f_+)c_-^2\frac{|S|^2|\phi_-|^2}{\Mpl^4}
-(1+f_-)c_+^2\frac{|S|^2|\phi_+|^2}{\Mpl^4}~.
\end{eqnarray}
Using Eqs.~(\ref{potenFterms}) and (\ref{Kmetricinvers}) we
calculate the F-term of the scalar potential:
\begin{eqnarray}
\label{Fnew}
V_F&=&\lambda^2
\frac{e^{K/\Mpl^2}}{g}\Biggl\{(1+f_+)(1+f_-)|\phi_+\phi_-|^2
\left[1+\frac{|S|^2}{\Mpl^2}(1+\Delta)\right]^2\nonumber\\
&&~~~~~~~~~~~~+(1+\Delta)(1+f_-)|S\phi_-|^2\left[1+\frac{|\phi_+|^2}{\Mpl^2}(1+f_+)
\right]^2\nonumber\\
&&~~~~~~~~~~~~-c_-^2\frac{|S\phi_-|^4}{\Mpl^4}\left[1+\frac{|\phi_+|^2}{\Mpl^2}(1+f_+)
\right]^2\nonumber\\
&&~~~~~~~~~~~~+(1+\Delta)(1+f_+)|S\phi_+|^2\left[1+\frac{|\phi_-|^2}{\Mpl^2}(1+f_-)\right]^2
\nonumber\\
&&~~~~~~~~~~~~-c_+^2\frac{|S\phi_+|^4}{\Mpl^4}\left[1+\frac{|\phi_-|^2}{\Mpl^2}(1+f_-)
\right]^2 \nonumber\\
&&~~~~~~~~~~~~-2c_+(1+f_+)\frac{|S\phi_-\phi_+|^2}{\Mpl^2}
\nonumber\\
&&~~~~~~~~~~~~~~~~\times
\left[1+\frac{|S|^2}{\Mpl^2}(1+\Delta)\right]
\left[1+\frac{|\phi_+|^2}{\Mpl^2}(1+f_+)\right]\nonumber \\
&&~~~~~~~~~~~~-2c_-(1+f_-)\frac{|S\phi_-\phi_+|^2}{\Mpl^2}
\nonumber\\
&&~~~~~~~~~~~~~~~~\times
\left[1+\frac{|S|^2}{\Mpl^2}(1+\Delta)\right]
\left[1+\frac{|\phi_-|^2}{\Mpl^2}(1+f_-)\right]\nonumber\\
&&~~~~~~~~~~~~+2c_+c_-|\phi_+\phi_-|^2\frac{|S|^4}{\Mpl^4}
\nonumber\\
&&~~~~~~~~~~~~~~~~\times
\left[1+\frac{|\phi_+|^2}{\Mpl^2}(1+f_+)\right]
\left[1+\frac{|\phi_-|^2}{\Mpl^2}(1+f_-)\right]\Biggr\}\nonumber\\
&&-3e^{K/\Mpl^2}\frac{|W|^2}{\Mpl^2}~.
\end{eqnarray}
From the F-term part of the scalar potential, different limits can be
taken to recover known results. For example, the limit $c_+=c_-=b=0$
allows to recover the standard minimal D-term inflation as discussed
in Ref.~\cite{DSUGRA2}. The limit $b=0$, leads to the potential
analysed in Section~\ref{secpotenseto} and first studied in
Ref.~\cite{seto}. Finally, the limit $c_+=c_-=0$, gives the case of
the scalar potential studied also in Ref.~\cite{seto} with the choice
$h(|S|^2)=|S|^2+|S|^4$.

\begin{figure}[hhh]
\begin{center}
\includegraphics[scale=.5]{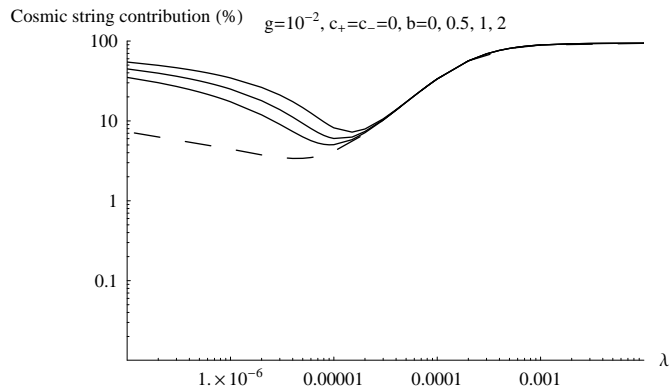}
\caption{Cosmic string contribution to the CMB temperature anisotropies as a
function of the superpotential coupling constant $\lambda$, in the
next-to-minimal D-term inflationary model. The value of the gauge
coupling constant is fixed and equal to $g=10^{-2}$; there are no
cross terms between $S$ and $\Phi_\pm$, i.e., $c_\pm=0$. The minimal
SUGRA for $b=0$ is represented by the dashed line, while the different
plain lines are calculated for $b=0.5,1,2$, going from the bottom to
the top. A fine tuning of the superpotential coupling $\lambda$ is
still required, in order to avoid a dominant contribution of cosmic
strings.}
\label{CSdelambdaNONMINEXTc0g2}
\end{center}
\end{figure}

\begin{figure}[hhh]
\begin{center}
\includegraphics[scale=.5]{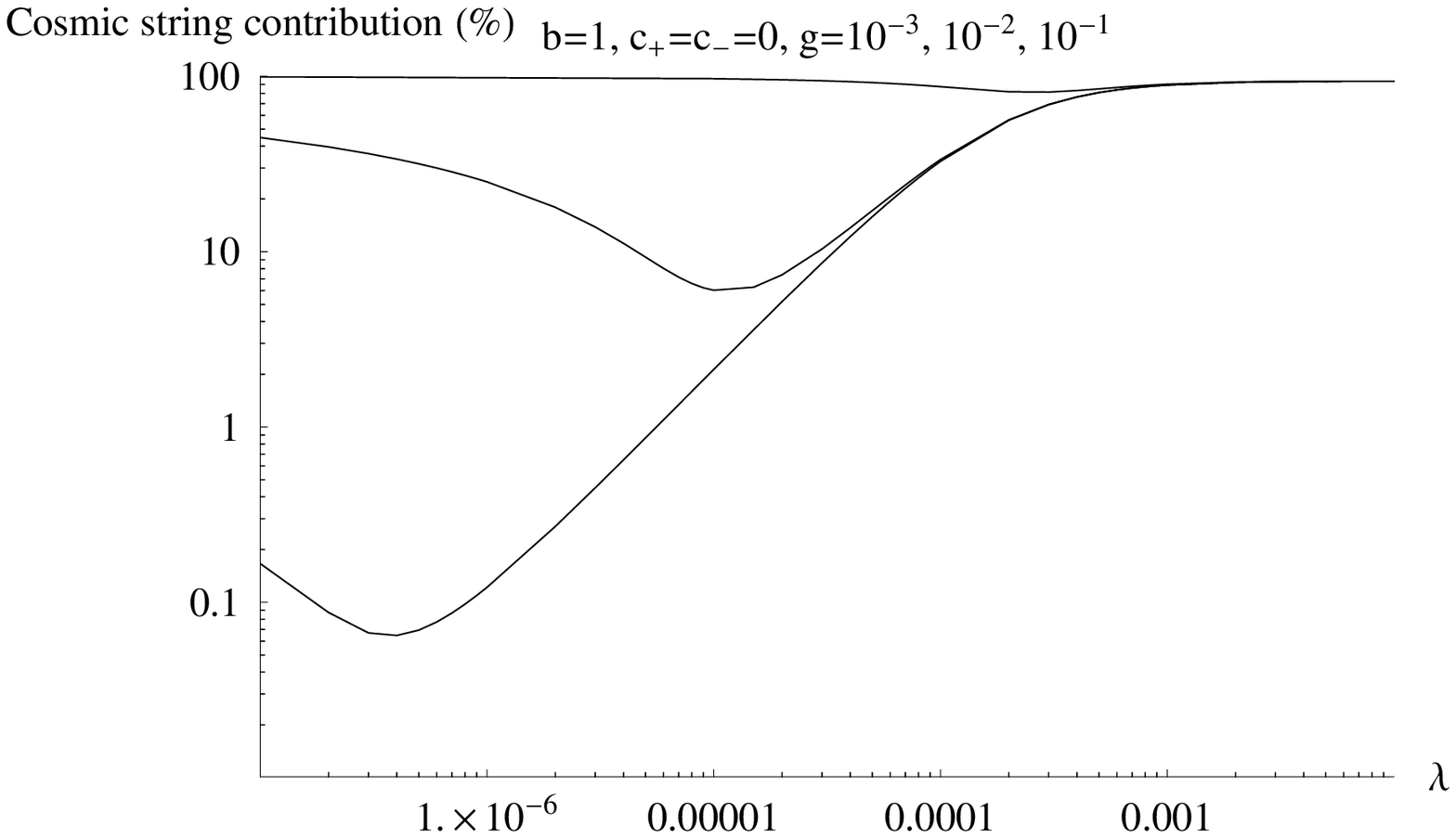}
\caption{Cosmic string contribution to the CMB temperature
anisotropies as a function of the superpotential coupling constant
$\lambda$, in the next-to-minimal D-term inflation model. The
parameter $b$ is set equal to $b=1$; there are no cross terms between
$S$ and $\Phi_\pm$, ie $c_\pm=0$. The different plain lines are
calculated for $g=10^{-3},10^{-2},10^{-1}$, going from the bottom to
the top. A fine tuning of the superpotential coupling $\lambda$ is
still is required to avoid a dominant contribution of cosmic strings.}
\label{CSdelambdaNONMINEXTc0b1}
\end{center}
\end{figure}

\begin{figure}[hhh]
\begin{center}
\includegraphics[scale=.5]{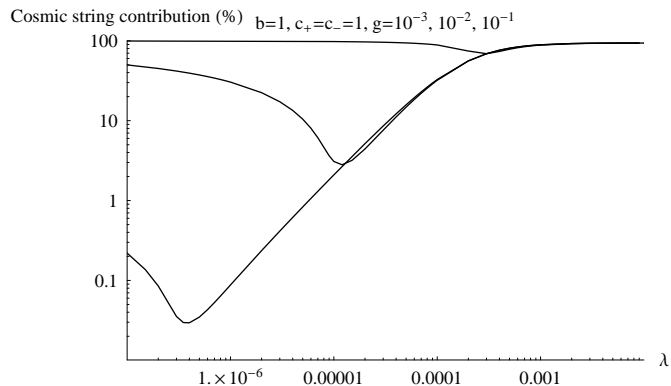}
\caption{Cosmic string contribution to the CMB temperature
anisotropies as a function of the superpotential coupling constant
$\lambda$, in the next-to-minimal D-term inflation model. The
parameter $b$ is set equal to $b=1$; there are no cross terms between
$S$ and $\Phi_\pm$, i.e., $c_\pm=0$. The different plain lines are
calculated for $g=10^{-3},10^{-2},10^{-1}$, going from the bottom to
the top.  A fine tuning of the superpotential coupling $\lambda$ is
still required, in order to avoid a dominant contribution of cosmic
strings.} \label{CSdelambdaNONMINEXTb1c1}
\end{center}
\end{figure}

Equation (\ref{Fnew}) implies that during inflation $V_F$ is minimised
for $\langle\phi_+\rangle=\langle\phi_+\rangle=0$, thus $V_F=0$ and
inflation is driven by the D-term. The total scalar potential $V$ is
constant at tree level and equal to $V_0=g^2\xi^2/2$. Having the
scalar potential we calculate the masses of the
canonically normalised scalar components of the superfields
$\Phi_\pm$:
\begin{equation}
m^2_\pm= \exp\left(\frac{|S|^2}{\Mpl^2}+b\frac{|S|^4}{\Mpl^4}\right)
\frac{\lambda^2|S|^2}{(1+f_+)(1+f_-)}+g^2q_\pm\xi~,
\end{equation}
whereas from Eq.~(\ref{lagrangfermio}) we get the mass squared for their
fermionic partners:
\begin{equation}
m^2_{\rm fermion}= \exp\left(\frac{|S|^2}{\Mpl^2}+b\frac{|S|^4}{\Mpl^4}\right)
\frac{\lambda^2|S|^2}{(1+f_+)(1+f_-)}~.
\end{equation}
Following the same procedure as in the previous sections, we obtain the
effective potential
\begin{equation}
V_{\rm eff}(|S|)=\frac{g^2\xi^2}{2}\left\{
1+\frac{g^2}{16\pi^2}\left[ 2\ln \left(
z\frac{g^2\xi}{\Lambda^2}\right)+f_V(z) \right] \right\} ~,
\end{equation}
where $f_V(z)$ is the same as in Section~\ref{secpotenseto},
Eq.~(\ref{deffV}); the only difference is the expression for $z$,
namely here
\begin{equation}
z\equiv \frac{\lambda^2|S|^2}{g^2\xi}\exp\bigg(\frac{|S|^2}{\Mpl^2}
+b\frac{|S|^4}{\Mpl^4}\bigg) \frac{1}{(1+f_+)(1+f_-)}~.\\
\end{equation}
The first derivative of the scalar potential with respect to $|S|$
reads
\begin{equation}
V'_{\rm eff}(|S|)\equiv\frac{dV_{\rm eff}}{d|S|}
=\frac{g^4\xi^2}{16\pi^2}\,z\,f_{V'}(z)\,f_z(|S|)~,
\end{equation}
where $f_{V'}$ is the same as in Section~\ref{secpotenseto},
 Eq.~(\ref{deffV'}), but $f_z(|S|)$ is in this case given by
\begin{equation}
f_z(|S|)=2|S|\left[ \frac{1}{\Mpl^2}+\frac{2b|S|^2}{\Mpl^4}
+\frac{1}{|S|^2}-\frac{c_+}{(1+f_+)\Mpl^2}-\frac{c_-}{(1+f_-)\Mpl^2}\right]~.
\label{newfz}
\end{equation}
The number of e-folds and the inflationary contribution to the CMB
quadrupole anisotropy are still calculated using
Eqs.~(\ref{efoldings}) and (\ref{dTsurT}), respectively, with
$f_z(|S|)$ given by Eq.~(\ref{newfz}) above. The cosmic strings
contribution to the CMB temperature anisotropies is the same as in
the minimal D-term SUGRA case. The contribution of cosmic strings
to the CMB anisotropies in the whole parameters space is
represented in Figs.~\ref{CSdelambdaNONMINEXTc0g2},
\ref{CSdelambdaNONMINEXTc0b1}, \ref{CSdelambdaNONMINEXTb1c1}.

Studying the above figures one easily concludes that considering a
more general form for the K\"ahler potential does not solve the fine
tuning problem of D-term inflation. The new term in the K\"ahler
potential, whose weight is given by the parameter $b$, induces an
enhancement of the cosmic string contribution at low $\lambda$.  We
still observe a dominant contribution of cosmic strings if the
superpotential coupling $\lambda$ is close to unity.  Therefore, the
constraints on $\lambda$ found on the previous section remain
unchanged as given in Eq.~(\ref{contraintelambdanonmin}).

\section*{Conclusions}

D-term hybrid inflation is a successful and interesting model.  In
the context of supergravity, D-term inflation avoids the {\sl
Hubble-induced mass} problem, which plagues F-term hybrid
inflation, while it can easily be implemented in string theory. In
a standard formulation of D-term inflation, where the constant FI
term gets compensated by a single complex scalar field, we do not
add an additional discrete symmetry, and do not consider
non-renormalisable terms in the potential, D-term strings are
formed at the end of the phase transitions which signals the end
of inflation. These strings are analogous to the D-strings formed
at the end of brane inflation, which is the result of brane
collisions.

D-term inflation cannot be studied in the standard formulation of
supergravity, which is ill-defined whenever the superpotential
vanishes and there are present constant Fayet Iliopoulos terms.
Following an effective supergravity formulation based on
superconformal theory, the superpotential transforms under the
U(1) symmetry along the directions where the FI terms are
constant. This transformation defines the charge assignments of
the superfields. D-term inflation has to be studied within this
new formulation of supergravity, which is well defined when the
superpotential vanishes.

Cosmic strings contribute to the cosmic microwave background
temperature anisotropies. All current measurements put severe
constraints on the allowed cosmic string contribution. To achieve
a compatibility between measurements and theoretical predictions
one should fine tune the couplings. This was already found in the
case of minimal supergravity. It was therefore natural to ask the
question of whether this result still holds in non-minimal
supergravity. It was previously claimed in the literature that
higher order K\"ahler potentials suppress the cosmic strings
contribution.  Studying a case of non-minimal supergravity, where
we include higher order corrections in the K\"ahler potential, as
well as supergravity with shift symmetry we conclude that cosmic
string contribution will be dominant unless the couplings are
fine-tuned. We also find that, as in the minimal
case~\cite{jcap2005,prl2005}, the $9\%$ constraint on the cosmic
string contribution is equivalent to the constraint
\begin{equation}
\sqrt{\xi}\leq 2.2\times 10^{15} \;\mathrm{GeV}
\Longleftrightarrow G\mu \leq 8.4\times 10^{-7}~.
\end{equation}
We would also like to emphasise that if $\sqrt{\xi}$ is higher by
a factor of 2, the cosmic string contributison is of the order of
$100\%$, as show in Fig.~7 of Ref.~\cite{jcap2005}.

In conclusion, we definitely disagree with the statement that
non-minimal K\"ahler potentials avoid the cosmic strings problem,
which should imply that fine tuning is not necessary. Even though we
have not studied a large number of K\"ahler potentials, our current
findings indicate that the problem of fine tuning is unavoidable
unless one considers a more complicated model, for example where
strings become topologically unstable, namely semi-local strings. We
have not yet studied~\cite{inprep} the spectral index in our models to
check the consistency with the limits imposed by the recent three-year
WMAP data, which however have been given only for purely adiabatic
models.  If the spectral index is higher than the one preferred from
the measurements and since, in addition, the requirement for very
small couplings seems more difficult to be satisfied in string theory,
one should then look for mechanisms to suppress the r\^ole of strings,
for example by making them unstable, along the lines of
Refs.~\cite{toine1,uad}, by adding new terms in the
superpotential~\cite{mcdonald}, or by considering the curvaton
mechanism~\cite{endoetal,prl2005}.

As we were completing this work, Ref.~\cite{spind} came to our
attention.  In that study, the authors find, in the context of minimal
supergravity, slightly higher upper bounds for the parameters
$\lambda$ and $g$, whereas the obtain the same with us upper bound on
the FI term. This small discrepancy originates from the different
analysis followed in Ref.~\cite{spind}; the authors perform a full
Markov-Chain-Monte-Carlo analysis and consider one more parameter,
namely the spectral index.

\section*{Acknowledgments}

It is a pleasure to thank N.\ Chatillon, R.\ Kallosh, P.\ Greene,
J.\ Gray, J.\ Martin, A.\ Van Proeyen and J.\ Yokoyama for
stimulating discussions. JR acknowledges support by the National Science
Foundation under Grant No. PHY-0455649 and by the University of Texas at San Antonio,
Texas 78249, USA.

\end{document}